\def\be{\begin{equation}}

\def\ee{\end{equation}}
\def\ba{\begin{array}}
\def\ea{\end{array}}

\def\Nb{{I\!\! N}}

\documentclass[12pt]{article}
\usepackage{amssymb}
\begin{document}

\setcounter{page}{1}
\centerline{\Large\bf Hermitian Tensor Product
Approximation} \vspace{2ex}
\centerline{\Large\bf  of Complex Matrices and Separability}
\vspace{4ex}
\begin{center}

Shao-Ming Fei$^{a,b}$, ~Naihuan Jing$^{c, d}$, ~Bao-Zhi Sun$^{a}$

\vspace{2ex}
\begin{minipage}{5in}

{\small $~^{a}$ Department of Mathematics, Capital Normal
University, Beijing 100037}

{\small $~^{b}$ Max-Planck-Institute for Mathematics in the Sciences, 04103 Leipzig}

{\small $~^{c}$ Department of Mathematics, North Carolina State University,

~~~Raleigh, NC 27695}

{\small $~^{d}$ Department of Mathematics,
Hubei University, Wuhan, Hubei 430062}

\end{minipage}
\end{center}

\vskip 1 true cm
\parindent=18pt
\parskip=6pt
\begin{center}
\begin{minipage}{4.5in}
\vspace{3ex} \centerline{\large Abstract} \vspace{4ex}

The approximation of matrices to the sum of tensor products of
Hermitian matrices is studied. A minimum decomposition of
matrices on tensor space $H_1\otimes H_2$ in terms of the sum of tensor
products of Hermitian matrices on $H_1$ and $H_2$ is presented.
From this construction the separability of quantum states
is discussed.
\bigskip
\bigskip

PACS numbers: 03.67.-a, 03.65.Ud, 03.65.Ta\vfill
\smallskip
MSC numbers: \vfill
\smallskip

Key words: Separability; Matrix; Tensor product
decomposition\vfill

\end{minipage}
\end{center}
\bigskip
\bigskip

\section{Introduction}

The quantum entangled states have become one of the
key resources in quantum information processing.
The study of quantum teleportation, quantum
cryptography, quantum dense coding, quantum error correction
and parallel computation \cite{pre98,nielsen,zeilinger} has
spurred a flurry of activities in the investigation of quantum
entanglements. Despite the potential applications of quantum entangled
states, the theory of quantum entanglement itself is far from
being satisfied. The separability for bipartite
and multipartite quantum mixed states is one of the
important problems in quantum entanglement.

Let $H_1$ (resp. $H_2$) be an
$m$ (resp. $n$)-dimensional complex Hilbert space, with $\vert i\rangle$, $i=1,...,m$
(resp. $\vert j\rangle$, $j=1,...,n$),
as an orthonormal basis.
A bipartite mixed state is said to be separable if the density matrix can be written as
\begin{equation}\label{seprho}
\rho=\sum_{i}p_{i}\rho _{i}^{1}\otimes \rho _{i}^{2},  \label{sep}
\end{equation}
where $0< p_i\leq 1$, $\sum_i p_i =1$,
$\rho _{i}^{1}$ and $\rho _{i}^{2}$ are rank one density matrices
on $H_1$ and $H_2$ respectively.
It is a challenge to find a decomposition like (\ref{sep}) or proving that it
does not exist for a generic mixed state $\rho$
\cite{lbck00,terhal01,3hreview}. With
considerable effort in analyzing the separability,
there have been some (necessary) criteria for separability in recent years, for instance,
the Bell inequalities \cite{Bell64},
PPT (positive partial transposition)\cite{peres} (which is also sufficient for
the cases $2\times 2$ and $2\times 3$ bipartite systems \cite{3hPLA223}),
reduction criterion\cite{2hPRA99,cag99}, majorization criterion\cite{nielson01},
entanglement witnesses \cite{3hPLA223} and
\cite{ter00,lkch00}, realignment \cite{ru02,ChenQIC03,chenPLA02} and
generalized realignment \cite{chenkai}, as well as
some necessary and sufficient criteria
for low rank density matrices \cite{hlvc00,afg01,feipla02} and also
for general ones but not operational \cite{3hPLA223}.

In \cite{loan} the minimum distance (in the sense of matrix norm)
between a given matrix and some other matrices with certain rank
is studied. In \cite{loan1} and \cite{kropro}, for a given matrix
$A$, the minimum of the Frobenius norm like
$||A-\sum_{i}B_i\otimes C_i||_F$ is investigated.
In this paper we develop the method of Hermitian
tensor product approximation for general complex matrix $A$, i.e.
we require $B_i$ and $C_i$ to be Hermitian matrices. By dealing with
the Hermitian condition as higher dimensional real constraints, an
explicit construction of general matrices on $H_1\otimes H_2$
according to the sum of the tensor products of Hermitian matrices
as well as real symmetric matrices on $H_1\otimes H_2$ is
presented. The results are generalized to the multipartite case.
The separability problem is discussed in terms of these tensor
product expressions.

\section{Tensor product decomposition in terms of real symmetric matrices}

We first consider the tensor product decompositions according to real symmetric matrices.
Let $A$ be a given $mn\times mn$ real matrix on $H_1\otimes H_2$.
We consider the problem of approximation of $A$ such that the Frobenius norm
\be\label{minr}
||A-\sum_{i}^r B_i\otimes C_i||_F
\ee
is minimized for some $m\times m$
real symmetric matrix $B_i$ on $H_1$ and $n\times n$ real symmetric matrix $C_i$ on $H_2$,
$i=1,...,r\in \Nb$.

We first introduce some notations.
For an $m\times m$ block matrix $Z$ with each block $Z_{ij}$ of size
$n\times n$, $i,j=1,...,m$, the realigned matrix $\tilde{Z}$ is defined by
$$
\tilde{Z}=[vec(Z_{11}),\cdots,vec(Z_{m1}),\cdots,
vec(Z_{1m}),\cdots, vec(Z_{mm})]^t,
$$
where for any $m\times n$ matrix $T$  with  entries $t_{ij}$,
$vec(T)$ is defined to be
$$
vec(T)=[t_{11},\cdots,t_{m1},t_{12},\cdots,t_{m2},\cdots,t_{1n},\cdots,
t_{mn}]^t.
$$
There is also another useful definition of $\tilde{Z}$,
$(\tilde{Z})_{ij,kl}=(Z)_{ik,jl}$.
A matrix $Z$ can
be expressed as the tensor product of two matrices
$V_1$ on $H_1$ and $V_2$ on $H_2$, $Z=V_1\otimes V_2$ if and only if
(cf, e.g., \cite{kropro})
$\tilde{Z}=vec(V_1)vec(V_2)^t$, i.e., the rank of
$\tilde{Z}$ is one,  $r(\tilde{Z})=1$.

Due to the property of the Frobenius norm, we have
\be\label{abcr}
||A-\sum_{i=1}^r B_i\otimes C_i||_F=||\tilde{A}-\sum_{i=1}^r vec(B_i)vec(C_i)^t||_F.
\ee
The symmetric condition of the matrices $B_i$ and $C_i$
can be expressed in terms of some real matrices $S_1$ and $S_2$ in a form
\be\label{condr}
S_1^t\, vec(B_i)=S_2^t\, vec(C_i)=0,~~~~i=1,...,r.
\ee

We define $Q_s$ to be an $m^2\times\frac{m(m-1)}{2}$ matrix such that, if we arrange the row
indices of $Q_s$ as $\{11,21,31,...,m1, 12,22,32,...,m2,...,mm\}$, then all the
entries of $Q_s$ are zero except those at $21$ and $12$ (resp. $31$ and
$13$, ...) which are $1$
and $-1$ respectively in the first (resp. second, ...) column.
We simply denote
\be\label{qs}
Q_s=[\{e_{21},-e_{12}\};\{e_{31},-e_{13}\};...;\{e_{m,m-1},-e_{m-1,m}\}],
\ee
where $\{e_{21},-e_{12}\}$ is first column of $Q_s$, with $1$ and $-1$ at the
$21$ and $12$ rows respectively; while
$\{e_{31},-e_{13}\}$ is second column of $Q_s$, with $1$ and $-1$ at the
$31$ and $13$ rows respectively; and so on.

Similarly we define $Q_a$ to be an $m^2\times\frac{m(m+1)}{2}$ matrix such that
\be\label{qa}
Q_a=[\{e_{11}\};\{e_{21},e_{12}\};\{e_{31},e_{13}\};...;\{e_{22}\};\\ \{e_{32},e_{23}\};
\{e_{42},e_{24}\};...;\{e_{m,m-1},e_{m-1,m}\},\{e_{mm}\}].
\ee
For $m=2$, we have
$$
Q_s=\left(\ba{c}0\\1\\-1\\0\ea\right),~~~~~
Q_a=\left(\ba{ccc}1&0&0\\0&1&0\\0&1&0\\0&0&1\ea\right).
$$

$S_1$ can then be expressed as, something like QR decomposition,
\be\label{QRr}
S_1=Q_s\equiv Q_1\left(\ba{c}R_1\\0\ea\right),
\ee
where $R_1$ is a full rank $\frac{m(m-1)}{2}\times \frac{m(m-1)}{2}$ matrix, $Q_1$ is an orthogonal
matrix, $Q_1=\left(\bar{Q}_s\bar{Q}_a\right)$, where $\bar{Q}_s$ and $\bar{Q}_a$
are obtained by normalizing the norm
of every column vector of $Q_s$ and $Q_a$ to be one.

For the case $m=2$,
\be\label{xyr2r}
Q_1=\left(\ba{cccc}0&1&0&0\\[2mm]
\frac{1}{\sqrt{2}}&0&\frac{1}{\sqrt{2}}&0\\[2mm]
-\frac{1}{\sqrt{2}}&0&\frac{1}{\sqrt{2}}&0\\[2mm]
0&0&0&1\ea\right),~~~
R_1=\left(\sqrt{2}\right).
\ee

$S_2$ has a similar QR decomposition with
$S_2=Q_2\left(\ba{c}R_2\\0\ea\right)$,
by replacing the dimension $m$ with $n$ in (\ref{QRr}).

Set
\be\label{setr}
 Q_1^t\tilde{A}Q_2=
\left(\ba{cc}\hat{A}_{11}&\hat{A}_{12}\\\hat{A}_{21}&\hat{A}_{22}
\ea\right).
\ee
Suppose the singular value decomposition of
$\hat{A}_{22}$ is given by $\hat{A}_{22}=\sum_{i=1}^r\sqrt{\lambda_i} u_i
v_i^t$, where $r$ and $\lambda_i$, $i=1,2,...,r$, are the rank and
eigenvalues of $\hat{A}_{22}^\dag\hat{A}_{22}$ respectively and
$u_i$ (resp. $v_i$) are the eigenvectors of the matrix
$\hat{A}_{22}\hat{A}_{22}^\dag$ (resp.
$\hat{A}_{22}^\dag\hat{A}_{22}$). Set
$\hat{\mathcal{B}}_i=\sqrt{\lambda_i} u_i$, $\hat{\mathcal{C}}_i=v_i$.

{\sf [Theorem 1]}.
Let $A$ be an $mn\times mn$ real matrix on $H_1\otimes H_2$, where
$dim(H_1)=m$, $dim(H_2)=n$. The minimum of the Frobenius norm
$||A-\sum_i^r B_i\otimes C_i||_F$ is obtained for some $m\times m$
real symmetric matrix $B_i$ on $H_1$ and $n\times n$ real symmetric matrix $C_i$ on $H_2$,
given by
\be\label{set2r}
vec(B_i)
=Q_1\left(\ba{c}0\\\hat{{\mathcal{B}}_i}\ea\right),~~~~
vec(C_i)
=Q_2\left(\ba{c}0\\\hat{{\mathcal{C}}_i}\ea\right).
\ee

{\sf [Proof].} Set \be\label{set1r} \ba{l} Q_1^t\, vec(B_i)
=\left(\ba{c}\hat{\textsc{b}}_i\\\hat{{\mathcal{B}}_i}\ea\right),~~~~
Q_2^t\, vec(C_i)
=\left(\ba{c}\hat{\textsc{c}}_i\\\hat{{\mathcal{C}}_i}\ea\right).
\ea \ee From (\ref{condr}) and (\ref{QRr}) we have
$$
\left(\ba{cc}R_1^t&0\ea\right)\left(\ba{c}\hat{\textsc{b}}_i\\\hat{{\mathcal{B}}_i}\ea\right)=0,~~~
\left(\ba{cc}R_2^t&0\ea\right)\left(\ba{c}\hat{\textsc{c}}_i\\\hat{{\mathcal{C}}_i}\ea\right)=0,
$$
which give rise to $\hat{\textsc{b}}_i=\hat{\textsc{c}}_i=0$,
due to the nonsingularity of $R_1^t$ and $R_2^t$.

From (\ref{abcr}), (\ref{setr}) and (\ref{set1r}) we obtain
$$
\ba{l}
||A-\displaystyle\sum_{i=1}^r B_i\otimes C_i||_F=||Q^t_1\tilde{A}Q_2
-\displaystyle\sum_{i=1}^r Q^t_1 vec(B_i)vec(C_i)^tQ_2||_F\\[3mm]
=\left|\left|\left(\ba{cc}\hat{A}_{11}&\hat{A}_{12}\\\hat{A}_{21}&\hat{A}_{22}\ea\right)
-\displaystyle\sum_{i=1}^r\left(\ba{c}0\\\hat{\mathcal{B}}_i\ea\right)
\left(\ba{cc}0&\hat{\mathcal{C}}^t_i\ea\right)
\right|\right|_F\\[5mm]
=\left|\left|\left(\ba{cc}\hat{A}_{11}&\hat{A}_{12}\\\hat{A}_{21}&\hat{A}_{22}\ea\right)
-\displaystyle\sum_{i=1}^r\left(\ba{cc}0&0\\
0&\hat{\mathcal{B}}_i\hat{\mathcal{C}}^t_i\ea\right)
\right|\right|_F\\[5mm]
=\sqrt{||\hat{A}_{11}||_F^2+||\hat{A}_{12}||_F^2+||\hat{A}_{21}||_F^2+||\hat{A}_{22}
-\displaystyle\sum_{i=1}^r\hat{\mathcal{B}}_i\hat{\mathcal{C}}^t_i||_F^2}\,.
\ea
$$
From matrix approximation we have that
$\hat{A}_{22}=\sum_{i=1}^r\hat{\mathcal{B}}_i\hat{\mathcal{C}}^t_i$ is the singular
value decomposition (SVD) for $\hat{A}_{22}$, which results in formula (\ref{set2r}).
\hfill $\square$.

From Theorem 1 we see that if a real symmetric matrix $A$ has a decomposition of tensor
product of real symmetric matrices, then $\hat{A}_{11}=\hat{A}_{12}=\hat{A}_{21}=0$.
As an example we consider the Werner state \cite{werner},
\be\label{werner}
\rho_w=\frac{1-F}{3}I_{4\times 4}+\frac{4F-1}{3}|\Psi^-\rangle\langle\Psi^-|
=\left(\ba{cccc}\frac{1-F}{3}&0&0&0\\[1mm]
0&\frac{2F+1}{6}&\frac{1-4F}{6}&0\\[1mm]
0&\frac{1-4F}{6}&\frac{2F+1}{6}&0\\[1mm]
0&0&0&\frac{1-F}{3}\ea\right),
\ee
where $|\Psi^-\rangle=(|01\rangle-|10\rangle)/\sqrt{2}$. State
$\rho_w$ is separable for $F\leq 1/2$ and entangled for $1/2<F\leq 1$.
According to the definition of realignment we have
$$
\tilde{\rho}_w=\left(\ba{cccc}\frac{1-F}{3}&0&0&\frac{2F+1}{6}\\[1mm]
0&0&\frac{1-4F}{6}&0\\[1mm]
0&\frac{1-4F}{6}&0&0\\[1mm]
\frac{2F+1}{6}&0&0&\frac{1-F}{3}\ea\right).
$$
Here the dimension $m=n$, hence $Q_2=Q_1$
is given by (\ref{xyr2r}). From (\ref{setr}) we have
$$
Q_1^t\tilde{\rho}_wQ_2=
\left(\ba{cc}\hat{\rho}_{w11}&\hat{\rho}_{w12}\\\hat{\rho}_{w21}&\hat{\rho}_{w22}\ea\right)
=\left(\ba{cccc}\frac{4F-1}{6}&0&0&0\\[1mm]
0&\frac{1-F}{3}&0&\frac{2F+1}{6}\\[1mm]
0&0&\frac{1-4F}{6}&0\\[1mm]
0&\frac{2F+1}{6}&0&\frac{1-F}{3}\ea\right).
$$
Therefore $\rho_w$ is generally not decomposable according to real symmetric matrices because
$(\hat{\rho}_{w})_{11}=(4F-1)/6\neq 0$ as long as $F\neq 1/4$. From
the singular value decomposition of $(\hat{\rho}_{w})_{22}$,
$$
(\hat{\rho}_{w})_{22}=\left(\ba{ccc}\frac{1-F}{3}&0&\frac{2F+1}{6}\\[1mm]
0&\frac{1-4F}{6}&0\\[1mm]
\frac{2F+1}{6}&0&\frac{1-F}{3}\ea\right)
$$
we have:
$$
u_1=v_1=\frac{1}{\sqrt{2}}\left(\ba{c}1\\0\\1\ea\right),~~~
\varepsilon
u_2=v_2=\frac{1}{\sqrt{2}}\left(\ba{c}-1\\0\\1\ea\right),~~~
\varepsilon u_3=v_3=\left(\ba{c}0\\1\\0\ea\right),
$$
with eigenvalues $\lambda_1=1/4$,
$\lambda_2=\lambda_3=(1-4F)^2/36$ respectively, where
$\varepsilon=(1-4F)/|1-4F|$. From (\ref{set2r}) we have
$vec(B_1)=\sqrt{\lambda_1} (1/\sqrt{2},0,0,1/\sqrt{2})^t$,
$vec(B_2)=\varepsilon\sqrt{\lambda_2}
(-1/\sqrt{2},0,0,1/\sqrt{2})^t$,
$vec(B_3)=\varepsilon\sqrt{\lambda_3}
(0,1/\sqrt{2},1/\sqrt{2},0)^t$. Therefore the best real symmetric
matrix tensor product decomposition is
$$
\rho_w\approx
\frac{1}{4}I_{2\times 2}\otimes I_{2\times 2}
+\frac{1-4F}{12}(\sigma_1\otimes \sigma_1+\sigma_3\otimes \sigma_3)\,,
$$
where $\sigma_i$ are Pauli matrices
$\sigma_1=\left(\ba{cc}0&1\\1&0\ea\right)$,
$\sigma_2=\left(\ba{cc}0&-i\\i&0\ea\right)$,
$\sigma_3=\left(\ba{cc}1&0\\0&-1\ea\right)$.

\section{Hermitian tensor product decomposition of Hermitian matrices}

We consider now the tensor product decompositions according to Hermitian matrices.
Let $A$ be a given $mn\times mn$ complex matrix on
$H_1\otimes H_2$. We first consider the problem of approximation
of $A$ such that the Frobenius norm
\be\label{min} ||A-B\otimes C||_F=||\tilde{A}-vec(B)vec(C)^t||_F.
\ee
is minimized for some
$m\times m$ Hermitian matrix $B$ on $H_1$ and $n\times n$
Hermitian matrix $C$ on $H_2$.

In order to impose the Hermitian condition of the matrices $B$ and $C$,
we separate the matrices $B$ and $C$ into real and imaginary parts
such that $B=\textsc{b}+i\mathcal{B}$, $C=\textsc{c}+i\mathcal{C}$,
where $\textsc{b}$ and $\textsc{c}$ (resp. $\mathcal{B}$ and $\mathcal{C}$)
are the real (resp. imaginary) parts of $B$ and $C$
respectively. As $vec(B)=vec(\textsc{b})+i\,vec(\mathcal{B})$, we have
$$
vec(B)vec(C)^t=(vec(\textsc{b})vec(\textsc{c})^t-vec({\mathcal{B}})vec({\mathcal{C}})^t)+
i(vec({\textsc{b}})vec({\mathcal{C}})^t+vec({\mathcal{B}})vec({\textsc{c}})^t).
$$
We now map the complex matrix $A$ to be a real one:
$$
A~\longrightarrow~\left(\ba{cc}\textsc{a}&\mathcal{A}\\-\mathcal{A}&\textsc{a}\ea\right),
$$
where $\textsc{a}$ and $\mathcal{A}$
are the real and the imaginary parts of $A$ respectively.
Now the approximation problem of complex matrices to the tensor
product of two Hermitian matrices is reduced to the problem of
real matrices and the results in \cite{loan1,kropro} can be used accordingly.
The problem to minimize
$||\tilde{A}-vec(B)vec(C)^t||_F$ is reduced to minimize
\be\label{min1}
\left|\left|\left(\ba{cc}\tilde{\textsc{a}}&\tilde{\mathcal{A}}\\-\tilde{\mathcal{A}}&\tilde{\textsc{a}}\ea\right)
-\left(\ba{cc}vec(\textsc{b})&vec(\mathcal{B})\\-vec(\mathcal{B})&vec(\textsc{b})\ea\right)
\left(\ba{cc}vec(\textsc{c})&-vec(\mathcal{C})\\vec(\mathcal{C})&vec(\textsc{c})\ea\right)^t\right|\right|_F
\ee under the Hermitian condition: $B=B^\dag$, $C=C^\dag$, i.e.,
$\textsc{b}$ and $\textsc{c}$ are symmetric, $\mathcal{B}$ and
$\mathcal{C}$ are antisymmetric. This condition can be expressed
in terms of some real matrices $S_1$ and $S_2$ in a form
\be\label{cond} S_1^t\left(\ba{c}vec(\textsc{b})\\\pm
vec(\mathcal{B})\ea\right)= S_2^t\left(\ba{c}vec(\textsc{c})\\\pm
vec(\mathcal{C})\ea\right)=0. \ee

{\sf [Lemma 1].} Condition (\ref{cond}) has a QR decomposition such that
\be\label{QR}
S_1=Q_1\left(\ba{c}R_1\\0\ea\right),~~~S_2=Q_2\left(\ba{c}R_2\\0\ea\right),
\ee
where $R_1$ and $R_2$ are full rank matrices, $Q_1$ and $Q_2$ are orthogonal
matrices.

{\sf [Proof].} $S_1$ can be generally expressed as
$$
S_1=\left(\ba{cc}Q_s&0\\0&Q_a\ea\right),
$$
where $Q_s$ and $Q_a$ are given by (\ref{qs}) and (\ref{qa}) respectively.
The QR decomposition of $S_1$ is given by
\be\label{q1}
Q_1=\left(\ba{cccc}\bar{Q}_s&0&0&\bar{Q}_a\\0&\bar{Q}_a&\bar{Q}_s&0\ea\right)
\equiv\left(\ba{cc}X_1&Y_1\\Y_1&X_1\ea\right),
\ee
with $\bar{Q}_s$ and $\bar{Q}_a$ given in section 2, $X_1$ (resp. $Y_1$)
is an $m^2\times m^2$ matrix with the first $m(m-1)/2$ (resp. last $m(m+1)/2$)
columns replaced by the matrix $\bar{Q}_s$ (resp. $\bar{Q}_a$) and the rest entries zero,
$R_1$ is a diagonal matrix with diagonal elements either $1$ or $\sqrt{2}$.
For the case $m=2$,
\be\label{xyr2}
X_1=\left(\ba{cccc}0&0&0&0\\\frac{1}{\sqrt{2}}&0&0&0
\\-\frac{1}{\sqrt{2}}&0&0&0\\0&0&0&0\ea\right),~~~
Y_1=\left(\ba{cccc}0&1&0&0\\0&0&\frac{1}{\sqrt{2}}&0
\\0&0&\frac{1}{\sqrt{2}}&0\\0&0&0&1\ea\right),~~~
R_1=\left(\ba{cccc}\sqrt{2}&0&0&0\\0&1&0&0\\0&0&\sqrt{2}&0\\0&0&0&1\ea\right).
\ee

$S_2$ has a similar QR decomposition with
\be\label{q2}
Q_2=\left(\ba{cc}X_2&Y_2\\Y_2&X_2\ea\right),
\ee
by replacing the dimension $m$ with $n$ in the expression of $S_1$. \hfill $\square$

Set
\be\label{set1}
\ba{l}
Q_1^t\left(\ba{c}vec(\textsc{b})\\-vec(\mathcal{B})\ea\right)
\equiv\left(\ba{c}\hat{\textsc{b}}\\-\hat{{\mathcal{B}}}\ea\right),~~~~
Q_1^t\left(\ba{c}vec(\textsc{b})\\vec(\mathcal{B})\ea\right)
\equiv\left(\ba{c}\check{\textsc{b}}\\\check{{\mathcal{B}}}\ea\right),\\[5mm]
Q_2^t\left(\ba{c}vec(\textsc{c})\\-vec(\mathcal{C})\ea\right)
\equiv\left(\ba{c}\hat{\textsc{c}}\\-\hat{{\mathcal{C}}}\ea\right),~~~~
Q_2^t\left(\ba{c}vec(\textsc{c})\\vec(\mathcal{C})\ea\right)
\equiv\left(\ba{c}\check{\textsc{c}}\\\check{{\mathcal{C}}}\ea\right).
\ea
\ee
From (\ref{cond}) and (\ref{QR}) we have
$$
\ba{l}
\left(\ba{cc}R_1^t&0\ea\right)\left(\ba{c}\hat{\textsc{b}}\\-\hat{{\mathcal{B}}}\ea\right)=0,~~~
\left(\ba{cc}R_1^t&0\ea\right)\left(\ba{c}\check{\textsc{b}}\\\check{{\mathcal{B}}}\ea\right)=0,\\[3mm]
\left(\ba{cc}R_2^t&0\ea\right)\left(\ba{c}\hat{\textsc{c}}\\-\hat{{\mathcal{C}}}\ea\right)=0,~~~
\left(\ba{cc}R_2^t&0\ea\right)\left(\ba{c}\check{\textsc{c}}\\\check{{\mathcal{C}}}\ea\right)=0,
\ea
$$
which give rise to $\hat{\textsc{b}}=\check{\textsc{b}}=\hat{\textsc{c}}=\check{\textsc{c}}=0$,
due to the nonsingularity of $R_1^t$ and $R_2^t$.

Therefore we have
\be\label{pp}
\ba{l}
\left(\ba{c}vec(\textsc{b})\\-vec(\mathcal{B})\ea\right)
=Q_1\left(\ba{c}\hat{\textsc{b}}\\-\hat{{\mathcal{B}}}\ea\right)
=\left(\ba{cc}X_1&Y_1\\Y_1&X_1\ea\right)\left(\ba{c}0\\-\hat{{\mathcal{B}}}\ea\right)
=\left(\ba{c}-Y_1\hat{{\mathcal{B}}}\\-X_1\hat{{\mathcal{B}}}\ea\right),~~~~\\[4mm]
\left(\ba{c}vec(\textsc{b})\\vec(\mathcal{B})\ea\right)
=Q_1\left(\ba{c}\check{\textsc{b}}\\\check{{\mathcal{B}}}\ea\right)
=\left(\ba{cc}X_1&Y_1\\Y_1&X_1\ea\right)\left(\ba{c}0\\\check{{\mathcal{B}}}\ea\right)
=\left(\ba{c}Y_1\check{{\mathcal{B}}}\\X_1\check{{\mathcal{B}}}\ea\right).
\ea
\ee
Thus $-Y_1\hat{{\mathcal{B}}}=Y_1\check{{\mathcal{B}}}$, $X_1\hat{{\mathcal{B}}}=X_1\check{{\mathcal{B}}}$ and
\be\label{bb2}
\check{{\mathcal{B}}}=(X_1+Y_1)^{-1}(X_1-Y_1)\hat{{\mathcal{B}}}
=\left(\ba{c}\bar{Q}_s^t\\[4mm]\bar{Q}_a^t\ea\right)\left(\ba{cc}\bar{Q}_s&-\bar{Q}_a\ea\right)
\hat{{\mathcal{B}}}=I_{s,a}^m\hat{{\mathcal{B}}},
\ee
where $I_{s,a}^m=diag(I_s^m,-I_a^m)$, $I_s^m$ (resp. $I_a^m$) is an
$m(m-1)/2$ (resp. $m(m+1)/2$) dimensional identity matrix.

Let $P$ denote the permutation matrix,
$$
P=\left(\ba{cc}0&I_{m^2\times m^2}\\I_{m^2\times m^2}&0\ea\right).
$$
It is easily seen that $PQ_1P=Q_1$. From the second formula in (\ref{pp}) we have
$$
Q_1^t\left(\ba{c}vec(\mathcal{B})\\vec(\textsc{b})\ea\right)
=\left(\ba{c}\check{{\mathcal{B}}}\\0\ea\right).
$$
Hence we have \be\label{q1tb}
Q_1^t\left(\ba{cc}vec(\textsc{b})&vec(\mathcal{B})\\-vec(\mathcal{B})&vec(\textsc{b})\ea\right)
=\left(\ba{cc}0&\check{{\mathcal{B}}}\\-\hat{{\mathcal{B}}}&0\ea\right),
\ee and, similarly, \be\label{q2tc}
Q_2^t\left(\ba{cc}vec(\textsc{c})&-vec(\mathcal{C})\\
vec(\mathcal{C})&vec(\textsc{c})\ea\right)
=\left(\ba{cc}0&-\hat{{\mathcal{C}}}\\\check{{\mathcal{C}}}&0\ea\right),
\ee where \be\label{cc2}
\check{{\mathcal{C}}}=I_{s,a}^n\hat{{\mathcal{C}}}, \ee
$I_{s,a}^n=diag(I_s^n,-I_a^n)$, $I_s^n$ (resp. $I_a^n$) is an
$n(n-1)/2$ (resp. $n(n+1)/2$) dimensional identity matrix.

Set
\be\label{set}
Q_1^t\left(\ba{cc}\tilde{{\textsc{a}}}&\tilde{{\mathcal{A}}}\\
-\tilde{{\mathcal{A}}}&\tilde{{\textsc{a}}}\ea\right)Q_2
\equiv \left(\ba{cc}\hat{A}_{11}&\hat{A}_{12}\\\hat{A}_{21}&\hat{A}_{22}\ea\right).
\ee
That is
$$
\ba{rcl}
\hat{A}_{11}=X_1^t\tilde{{\textsc{a}}}X_2+Y_1^t\tilde{{\textsc{a}}}Y_2
+X_1^t\tilde{{\mathcal{A}}}Y_2-Y_1^t\tilde{{\mathcal{A}}}X_2,\\
\hat{A}_{12}=X_1^t\tilde{{\textsc{a}}}Y_2+Y_1^t\tilde{{\textsc{a}}}X_2
+X_1^t\tilde{{\mathcal{A}}}X_2-Y_1^t\tilde{{\mathcal{A}}}Y_2,\\
\hat{A}_{21}=Y_1^t\tilde{{\textsc{a}}}X_2+X_1^t\tilde{{\textsc{a}}}Y_2
+Y_1^t\tilde{{\mathcal{A}}}Y_2-X_1^t\tilde{{\mathcal{A}}}X_2,\\
\hat{A}_{22}=Y_1^t\tilde{{\textsc{a}}}Y_2+X_1^t\tilde{{\textsc{a}}}X_2
+Y_1^t\tilde{{\mathcal{A}}}X_2-X_1^t\tilde{{\mathcal{A}}}Y_2.
\ea
$$

\smallskip
{\sf [Theorem 2].} To minimize (\ref{min}) is equivalent to minimize
the following formula
\be\label{min2}
\sqrt{||\hat{A}_{22}
+\hat{\mathcal{B}}\check{\mathcal{C}}^t||_F^2+||\hat{A}_{11}
+\check{\mathcal{B}}\hat{\mathcal{C}}^t||_F^2+||\hat{A}_{12}||_F^2+||\hat{A}_{21}||_F^2}.
\ee

{\sf [Proof].}
From (\ref{min1}), (\ref{q1tb}) and (\ref{q2tc}) we obtain
$$
\ba{l}
\left|\left|\left(\ba{cc}\tilde{\textsc{a}}&\tilde{\mathcal{A}}\\-\tilde{\mathcal{A}}&\tilde{\textsc{a}}\ea\right)
-\left(\ba{cc}vec(\textsc{b})&vec(\mathcal{B})\\-vec(\mathcal{B})&vec(\textsc{b})\ea\right)
\left(\ba{cc}vec(\textsc{c})&-vec(\mathcal{C})\\ vec(\mathcal{C})&vec(\textsc{c})\ea\right)^t\right|\right|_F\\[5mm]
=\left|\left|Q_1^t\left(\ba{cc}\tilde{\textsc{a}}&\tilde{\mathcal{A}}\\-\tilde{\mathcal{A}}&\tilde{\textsc{a}}\ea\right)Q_2
-Q_1^t\left(\ba{cc}vec(\textsc{b})&vec(\mathcal{B})\\-vec(\mathcal{B})&vec(\textsc{b})\ea\right)
\left(\ba{cc}vec(\textsc{c})&-vec(\mathcal{C})\\ vec(\mathcal{C})&vec(\textsc{c})\ea\right)^t Q_2\right|\right|_F\\[5mm]
=\left|\left|\left(\ba{cc}\hat{A}_{11}&\hat{A}_{12}\\-\hat{A}_{21}&\hat{A}_{22}\ea\right)
-\left(\ba{cc}0&\check{\mathcal{B}}\\-\hat{\mathcal{B}}&0\ea\right)
\left(\ba{cc}0&-\hat{\mathcal{C}}\\\check{\mathcal{C}}&0\ea\right)^t
\right|\right|_F\\[5mm]
=\left|\left|\left(\ba{cc}\hat{A}_{11}&\hat{A}_{21}\\-\hat{A}_{21}&\hat{A}_{22}\ea\right)
-\left(\ba{cc}-\check{\mathcal{B}}\hat{\mathcal{C}}^t&0\\
0&-\hat{\mathcal{B}}\check{\mathcal{C}}^t\ea\right)
\right|\right|_F\\[5mm]
=\sqrt{||\hat{A}_{22}
+\hat{\mathcal{B}}\check{\mathcal{C}}^t||_F^2+||\hat{A}_{11}
+\check{\mathcal{B}}\hat{\mathcal{C}}^t||_F^2+||\hat{A}_{12}||_F^2+||\hat{A}_{21}||_F^2}\,. \ea
$$
\hfill $\square$

\smallskip
{\sf [Lemma 2].} If the matrix $A=\textsc{a}+i{\mathcal{A}}$ is Hermitian, we have the relations:
$$
\hat{A}_{12}=\hat{A}_{21}=0,~~~\hat{A}_{11}=I_{s,a}^m\hat{A}_{22}I_{s,a}^n.
$$

{\sf [Proof].}
As the matrix $A=\textsc{a}+i\mathcal{A}$ is Hermitian, i.e.,
$\textsc{a}$ is symmetric and $\mathcal{A}$ is antisymmetric, we have
$$
(\tilde{\textsc{a}})_{ij,kl}=(\textsc{a})_{ik,jl}=(\textsc{a})_{jl,ik}=(\tilde{\textsc{a}})_{ji,lk},~~~
(\tilde{\mathcal{A}})_{ij,kl}=({\mathcal{A}})_{ik,jl}=-({\mathcal{A}})_{jl,ik}=-(\tilde{{\mathcal{A}}})_{ji,lk}.
$$
Noting that in our construction of $X_{\alpha}$ and $Y_{\alpha}$,
$(X_{\alpha})_{ij,kl}=-(X_{\alpha})_{ji,kl}$, $(Y_{\alpha})_{ij,kl}=(Y_{\alpha})_{ji,kl}$,
$\alpha=1,2$, we obtain
$$
(X_1^t\tilde{\textsc{a}}Y_2)_{ij,pq}=(X_1)_{kl,ij}(\tilde{\textsc{a}})_{kl,mn}(Y_2)_{mn,pq}=
-(X_1^t\tilde{\textsc{a}}Y_2)_{ij,pq}=0.
$$
Similarly we have $Y_1^t\tilde{\textsc{a}}X_2=X_1^t\tilde{{\mathcal{A}}}X_2=Y_1^t\tilde{{\mathcal{A}}}Y_2=0$.
Hence $\hat{A}_{12}=\hat{A}_{21}=0$.

From the relations $I_{s,a}^mY_1=-Y_1$, $I_{s,a}^mX_1=X_1$ and
$Y_2I_{s,a}^n=-Y_2$, $I_{s,a}^nX_2=X_2$, we have
$I_{s,a}^m(Y_1^t\tilde{\textsc{a}}Y_2+X_1^t\tilde{\textsc{a}}X_2)I_{s,a}^n
=Y_1^t\tilde{\textsc{a}}Y_2+X_1^t\tilde{\textsc{a}}X_2$ and
$I_{s,a}^m(Y_1^t\tilde{{\mathcal{A}}}X_2-X_1^t\tilde{{\mathcal{A}}}Y_2)I_{s,a}^n
=-(Y_1^t\tilde{{\mathcal{A}}}X_2-X_1^t\tilde{{\mathcal{A}}}Y_2)$. Therefore
$\hat{A}_{11}=I_{s,a}^m\hat{A}_{22}I_{s,a}^n$ or $\hat{A}_{22}=I_{s,a}^m\hat{A}_{11}I_{s,a}^n$.
\hfill $\square$

For Hermitian matrix $A$, by using (\ref{bb2}), (\ref{cc2}) and Lemma 2 we have
$||\hat{A}_{11}+\check{\mathcal{B}}\hat{\mathcal{C}}^t||_F=
||\hat{A}_{11}+I_{s,a}^m\hat{\mathcal{B}}\check{\mathcal{C}}^tI_{s,a}^n||_F
=||I_{s,a}^m\hat{A}_{11}I_{s,a}^n+\hat{\mathcal{B}}\check{\mathcal{C}}^t||_F
=||\hat{A}_{22}+\hat{\mathcal{B}}\check{\mathcal{C}}^t||_F.$
From Lemma 2 we have that to
minimize $||A-B\otimes C||_F$ (\ref{min}) is equivalent to minimize
$||\hat{A}_{22} +\hat{\mathcal{B}}\check{\mathcal{C}}^t||_F$, which
maybe zero, when $\hat{A}_{22}=-\hat{\mathcal{B}}\check{\mathcal{C}}^t$.

Now the minimum of
the Frobenius norm $||A-\sum_i^r B_i\otimes C_i||_F$ can be obtained readily.
Suppose the singular value decomposition of
$\hat{A}_{22}$ is $\hat{A}_{22}=\sum_{i=1}^r\sqrt{\lambda_i} u_i
v_i^t$, where $r$ and $\lambda_i$, $i=1,2,...,r$, are the rank and
eigenvalues of $\hat{A}_{22}^\dag\hat{A}_{22}$ respectively and
$u_i$ (resp. $v_i$) are the eigenvectors of the matrix
$\hat{A}_{22}\hat{A}_{22}^\dag$ (resp.
$\hat{A}_{22}^\dag\hat{A}_{22}$). Set
$\hat{\mathcal{B}}_i=\sqrt{\lambda_i} u_i$,
$\check{\mathcal{C}}_i=-v_i$. By using the results in \cite{loan1,kropro},
for Hermitian matrix $A$
the minimum of $||A-\sum_{i=1}^r B_i\otimes C_i||_F$ is obtained when
$\hat{A}_{22}=-\sum_{i=1}^r\hat{\mathcal{B}}_i\check{\mathcal{C}}^t_i$.

{\sf [Theorem 3]}. Let $A$ be an $mn\times mn$ Hermitian matrix on
$H_1\otimes H_2$, where $dim(H_1)=m$, $dim(H_2)=n$. The minimum of
the Frobenius norm $||A-\sum_i^r B_i\otimes C_i||_F$ is obtained
for some $m\times m$ Hermitian matrix $B$ on $H_1$ and $n\times n$
Hermitian matrix $C$ on $H_2$, if
$\hat{A}_{22}=-\sum_{i=1}^r\hat{\mathcal{B}}_i\check{\mathcal{C}}^t_i$,
where $\hat{A}_{22}$ is defined by (\ref{set}),
$B_i={\textsc{b}}_i+i{\mathcal{B}}_i$,
$C_i={\textsc{c}}_i+i{\mathcal{C}}_i$, are given by the relations
\be\label{set2}
\left(\ba{c}vec(\textsc{b}_i)\\-vec({\mathcal{B}}_i)\ea\right)
=Q_1\left(\ba{c}0\\-\hat{{\mathcal{B}}_i}\ea\right),~~~~
\left(\ba{c}vec(\textsc{c}_i)\\ vec({\mathcal{C}}_i)\ea\right)
=Q_2\left(\ba{c}0\\ \check{{\mathcal{C}}_i}\ea\right). \ee

As an example we consider the bound entangled state on
$2\times 4$ ($m=2$, $n=4$) \cite{horodecki97}, \be\label{rhob}
\rho_b=\frac{1}{7b+1}\left( \ba{cccccccc}
b&0&0&0& 0&b&0&0\\
0&b&0&0& 0&0&b&0\\
0&0&b&0& 0&0&0&b\\
0&0&0&b& 0&0&0&0\\
0&0&0&0& \frac{1+b}{2}&0&0&\frac{\sqrt{1-b^2}}{2}\\
b&0&0&0& 0&b&0&0\\
0&b&0&0& 0&0&b&0\\
0&0&b&0& \frac{\sqrt{1-b^2}}{2}&0&0&\frac{1+b}{2}
\ea
\right),
\ee
where $0<b<1$. $\rho_b$ is a PPT but entangled state.
The QR decomposition in our case only depends the dimensions.
$Q_1$ is still given by (\ref{q1}) and (\ref{xyr2}).
$Q_2$ is a $32\times 32$ matrix with $X_2$, $Y_2$ in (\ref{q2})
given by $X_2=({\bf f}_1,{\bf f}_2,{\bf f}_3,{\bf f}_4,{\bf f}_5,{\bf f}_6,{\bf 0}_{10})$,
$Y_2=({\bf 0}_{6}, {\bf a}_1,{\bf a}_2,{\bf a}_3,{\bf a}_4,{\bf a}_5,
{\bf a}_6,{\bf a}_7,{\bf a}_8,{\bf a}_9,{\bf a}_{10})$, where ${\bf s}_i$
and ${\bf a}_i$ are $16\times 1$ column vectors:
$$\ba{l}
{\bf f}_1=(0,1/\sqrt{2},0,0, -1/\sqrt{2},0,0,0, 0,0,0,0, 0,0,0,0)^t,\\
{\bf f}_2=(0,0,1/\sqrt{2},0, 0,0,0,0, -1/\sqrt{2},0,0,0, 0,0,0,0)^t,\\
{\bf f}_3=(0,0,0,1/\sqrt{2}, 0,0,0,0, 0,0,0,0, -1/\sqrt{2},0,0,0)^t,\\
{\bf f}_4=(0,0,0,0, 0,0,1/\sqrt{2},0, 0,-1/\sqrt{2},0,0, 0,0,0,0)^t,\\
{\bf f}_5=(0,0,0,0, 0,0,0,1/\sqrt{2}, 0,0,0,0, 0,-1/\sqrt{2},0,0)^t,\\
{\bf f}_6=(0,0,0,0, 0,0,0,0, 0,0,0,1/\sqrt{2}, 0,0,-1/\sqrt{2},0)^t,\\
{\bf a}_1=(1,0,0,0, 0,0,0,0, 0,0,0,0, 0,0,0,0)^t,\\
{\bf a}_2=(0,1/\sqrt{2},0,0, 1/\sqrt{2},0,0,0, 0,0,0,0, 0,0,0,0)^t,\\
{\bf a}_3=(0,0,1/\sqrt{2},0, 0,0,0,0, 1/\sqrt{2},0,0,0, 0,0,0,0)^t,\\
{\bf a}_4=(0,0,0,1/\sqrt{2}, 0,0,0,0, 0,0,0,0, 1/\sqrt{2},0,0,0)^t,\\
{\bf a}_5=(0,0,0,0, 0,1,0,0, 0,0,0,0, 0,0,0,0)^t,\\
{\bf a}_6=(0,0,0,0, 0,0,1/\sqrt{2},0, 0,1/\sqrt{2},0,0, 0,0,0,0)^t,\\
{\bf a}_7=(0,0,0,0, 0,0,0,1/\sqrt{2}, 0,0,0,0, 0,1/\sqrt{2},0,0)^t,\\
{\bf a}_8=(0,0,0,0, 0,0,0,0, 0,0,1,0, 0,0,0,0)^t,\\
{\bf a}_9=(0,0,0,0, 0,0,0,0, 0,0,0,1/\sqrt{2}, 0,0,1/\sqrt{2},0)^t,\\
{\bf a}_{10}=(0,0,0,0, 0,0,0,0, 0,0,0,0, 0,0,0,1)^t,\\
\ea
$$
${\bf 0}_{6}$ and ${\bf 0}_{10}$ are $16\times 6$ and $16\times 10$
null matrices. From (\ref{set}) we have
$$
Q_1^t\left(\ba{cc}\tilde{\rho_b}&0\\
0&\tilde{\rho_b}\ea\right)Q_2
\equiv \left(\ba{cc}\hat{A}_{11}&0\\0&\hat{A}_{22}\ea\right),
$$
where
$$
\hat{A}_{11}=\hat{A}_{22}=\frac{1}{1+7b}\left(\ba{cccccccccccccccc}
b& 0& 0& b&     0& b& 0& 0&     0& 0& 0& 0&     0& 0& 0& 0\\
0& 0& 0& 0&      0& 0& b& 0&     0& 0& b& 0&    0& b& 0& b\\
0& 0& 0& 0&      0& 0& 0& b&     0& 0& 0& b&    0& 0& b& 0\\
0& 0& 0& 0&      0& 0& \frac{1 + b}{2}& 0&    0&
\sqrt{\frac{1 - b^2}{2}}& b& 0&       0& b& 0& \frac{1 + b}{2}
\ea
\right).
$$
From the singular value decomposition of $\hat{A}_{11}$ we have
$$
\ba{l l}
\hat{\mathcal{B}}_1=\frac{\sqrt{3}b}{1+7b}(0,0,1,0)^t,~~
&\hat{\mathcal{B}}_3=\frac{\sqrt{\lambda_-}}{(1+7b)\sqrt{1+D^2_+}}(0,D_+,0,1)^t,\\
\hat{\mathcal{B}}_2=\frac{\sqrt{3}b}{1+7b}(1,0,0,0)^t,~~
&\hat{\mathcal{B}}_4=\frac{\sqrt{\lambda_+}}{(1+7b)\sqrt{1+D^2_-}}(0,D_-,0,1)^t,
\ea
$$
and
$$
\ba{l}
\check{\mathcal{C}}_1=-\frac{1}{\sqrt{3}}(0,0,0,0, 0,0,0,1, 0,0,0,1, 0,0,1,0)^t,\\
\check{\mathcal{C}}_2=-\frac{1}{\sqrt{3}}(1,0,0,1, 0,1,0,0, 0,0,0,0, 0,0,0,0)^t,\\
\check{\mathcal{C}}_3=-\frac{1}{\sqrt{\lambda_-(1+D_+^2)}}(0,0,0,0,
0,0,bD_++\frac{1+b}{2},0, 0,
\sqrt{\frac{1-b^2}{2}},b(1+D_+),0, 0,b(1+D_+),0,bD_++\frac{1+b}{2})^t,\\
\check{\mathcal{C}}_4=-\frac{1}{\sqrt{\lambda_+(1+D_-^2)}}(0,0,0,0,
0,0,bD_-+\frac{1+b}{2},0, 0, \sqrt{\frac{1-b^2}{2}},b(1+D_-),0,
0,b(1+D_-),0,bD_-+\frac{1+b}{2})^t, \ea
$$
where
$$\lambda_\pm=\frac{1+b+6b^2\pm\sqrt{1+2b+b^2+20b^3+40b^4}}{2},$$
$$D_\pm=-\frac{1+b-2b^2\pm\sqrt{1+2b+b^2+20b^3+40b^4}}{2b(1+3b)}.$$
Using (\ref{set2}) we have
\be\label{rhobf} \ba{rcl}
\rho_b&=&\displaystyle\frac{b}{2(1+7b)}\left(\ba{cc}0&1\\1&0\ea\right)\otimes
\left(\ba{cccc}0&1&0&0\\1&0&1&0\\0&1&0&1\\0&0&1&0\ea\right)
+\displaystyle\frac{b}{2(1+7b)}\left(\ba{cc}0&-i\\i&0\ea\right)\otimes
\left(\ba{cccc}0&i&0&0\\-i&0&i&0\\0&-i&0&i\\0&0&-i&0\ea\right)\\[9mm]
&&+\displaystyle\frac{1}{(1+7b)(1+D_+^2)}\left(\ba{cc}D_+&0\\0&1\ea\right)\otimes
\left(\ba{cccc}\frac{1+b}{2}+bD_+&0&0&\frac{1-b^2}{2}\\0&b(1+D_+)&0&0\\
0&0&b(1+D_+)&0\\\frac{1-b^2}{2}&0&0&\frac{1+b}{2}+bD_+\ea\right)\\[12mm]
&&+\displaystyle\frac{1}{(1+7b)(1+D_-^2)}\left(\ba{cc}D_-&0\\0&1\ea\right)\otimes
\left(\ba{cccc}\frac{1+b}{2}+bD_-&0&0&\frac{1-b^2}{2}\\0&b(1+D_-)&0&0\\
0&0&b(1+D_-)&0\\\frac{1-b^2}{2}&0&0&\frac{1+b}{2}+bD_-\ea\right).
\ea
\ee

\section{Separability of bipartite mixed states}

In this section we discuss some properties related to the Hermitian
tensor product decomposition that could give rise to some
hints to the separability problem of bipartite mixed states.
From Theorem 3 we can always calculate the tensor product
decomposition in terms of Hermitian matrices for a given
density matrix $A$, $A=\sum_{i=1}^r B_i\otimes C_i$.
Nevertheless the Hermitian matrices $B_i$ and $C_i$ are
generally not positive. They are not density matrices defined
on the subspaces $H_1$ and $H_2$. Hence one can not say that
$A$ is separable.

Let $m(A)$ and $M(A)$ denote the smallest and the largest
eigenvalues of a Hermitian matrix $A$. We can transform the decomposition
into the one given by another set of
Hermitian matrices which all have the smallest eigenvalue zero, as follows,
\be\label{protocal}
\ba{rcl}
A&=&\displaystyle\sum_{i=1}^r B_i\otimes C_i=
\displaystyle\sum_{i=1}^r (B_i-m(B_i)I_m+m(B_i)I_m)\otimes (C_i-m(C_i)I_n+m(C_i)I_n)\\[5mm]
&=&\displaystyle\sum_{i=1}^r (B_i-m(B_i)I_m)\otimes (C_i-m(C_i)I_n)
+\displaystyle\sum_{i=1}^r m(C_i)(B_i-m(B_i)I_m)\otimes I_n\\[5mm]
&&+I_m\otimes \displaystyle\sum_{i=1}^r m(B_i)(C_i-m(C_i)I_n)
+\displaystyle\sum_{i=1}^r m(B_i)m(C_i)I_m\otimes I_n\\[5mm]
&=&\displaystyle\sum_{i=1}^r (B_i-m(B_i)I_m)\otimes (C_i-m(C_i)I_n)\\[5mm]
&&+\left[\displaystyle\sum_{i=1}^r m(C_i)(B_i-m(B_i)I_m)
-m\left(\displaystyle\sum_{i=1}^r m(C_i)(B_i-m(B_i)I_m)\right)I_m\right]\otimes I_n\\[5mm]
&&+I_m\otimes \left[\displaystyle\sum_{i=1}^r m(B_i)(C_i-m(C_i)I_n)
-m\left(\displaystyle\sum_{i=1}^r m(B_i)(C_i-m(C_i)I_n)\right)I_n\right]\\[5mm]
&&+\left[m\left(\displaystyle\sum_{i=1}^r m(C_i)B_i\right)
+ m\left(\displaystyle\sum_{i=1}^r m(B_i)C_i\right)
-\displaystyle\sum_{i=1}^r m(B_i)m(C_i)\right]I_m\otimes I_n\,,
\ea
\ee
where $I_m$ and $I_n$ stand for $m\times m$ and $n\times n$
identity matrices. The coefficient of $I_m\otimes I_n$,
$$
q_{B,C}\equiv m\left(\displaystyle\sum_{i=1}^r m(C_i)B_i\right)
+ m\left(\displaystyle\sum_{i=1}^r m(B_i)C_i\right)
-\displaystyle\sum_{i=1}^r m(B_i)m(C_i)
$$
associated with the decomposition $A=\sum_{i=1}^r B_i\otimes C_i$
is not necessary positive.

$q_{B,C}$ is decomposition dependent.
Associated with another decomposition $A=\sum_{i=1}^{r^\prime} B_i^\prime\otimes C_i^\prime$
one would obtain $q_{B^\prime,C^\prime}\neq q_{B,C}$. We define the
maximum value of $q_{B,C}$ to be the {\it separability indicator} of $A$, $S(A)=max(q_{B,C})$ for all possible
Hermitian decompositions of $A$. With respect to $S(A)$ the associated decomposition
is generally of the form
\be\label{sa}
A=\sum_{i} \bar{B}_i\otimes \bar{C}_i + I_m\otimes \bar{C} + \bar{B}\otimes I_n + S(A) I_m\otimes I_n,
\ee
where $\bar{B}_i\geq 0$, $\bar{C}_i\geq 0$ ,$\bar{B}\geq 0$, $\bar{C}\geq 0$ are positive Hermitian
matrices.

{\sf [Theorem 4].} Let $A$ be a Hermitian positive matrix with tensor product decompositions of Hermitian matrices
like $A=\sum_{i=1}^r B_i\otimes C_i$. $A$ is separable if and only if the separability indicator
$S(A)\geq 0$. Moreover $S(A)$ satisfies the following relations:
\be
S(A) \leq m(A),\label{thm3a}
\ee
\be\label{thm3b}
\ba{rcl}
S(A) &\geq& \displaystyle\frac{1}{2}\sum_{i=1}^r[M(B_i)m(C_i)+M(C_i)m(B_i)\\[4mm]
&&-|m(B_i)|(M(C_i)-m(C_i))-|m(C_i)|(M(B_i)-m(B_i))],
\ea
\ee

\be\label{thm3c}
S(A) \geq m(A)-\displaystyle\sum_{i} M(\bar{B}_i)M(\bar{C}_i).
\ee

{\sf [Proof].} If $A$ is separable, $A$ has a decomposition $A=\sum_{i} B_i\otimes C_i$
of the form (\ref{seprho}), i.e., $m(B_i)\geq 0$, $m(C_i)\geq 0$. We have
$$
\ba{rcl}
S(A)\geq q_{B,C}&=&
m\left(\displaystyle\sum_{i} m(C_i)B_i\right)
+ m\left(\displaystyle\sum_{i} m(B_i)C_i\right)
-\displaystyle\sum_{i} m(B_i)m(C_i)\\[5mm]
&\geq&\displaystyle\sum_{i} m(C_i)m(B_i)
+\displaystyle\sum_{i} m(B_i)m(C_i)
-\displaystyle\sum_{i} m(B_i)m(C_i)\\[5mm]
&=&\displaystyle\sum_{i} m(B_i)m(C_i)\geq 0\,.
\ea
$$
If $S(A)\geq 0$, then the associated decomposition (\ref{sa}) is already a
separable expression of $A$.

From the decomposition (\ref{sa}) with respect to $S(A)$, we have
$$
m(A)\geq \displaystyle\sum_{i} m(\bar{B}_i\otimes \bar{C}_i) +m(\bar{B}) +m(\bar{C}) +S(A)=S(A).
$$

On the other hand we have
$$\ba{rcl}
S(A)&\geq& q_{B,C}\geq
\displaystyle\sum_{i} m(m(C_i)B_i)
+ \displaystyle\sum_{i} m(m(B_i)C_i)-\displaystyle\sum_{i} m(B_i)m(C_i)\\[3mm]
&=&\displaystyle\sum_{i} \left( m(B_i)\frac{m(C_i)+|m(C_i)|}{2}
+M(B_i)\frac{m(C_i)-|m(C_i)|}{2}\right)\\[3mm]
&&+\displaystyle\sum_{i} \left( m(C_i)\frac{m(B_i)+|m(B_i)|}{2}
+M(C_i)\frac{m(B_i)-|m(B_i)|}{2}\right)-\displaystyle\sum_{i}m(B_i)m(C_i)\,,
\ea
$$
which is just the formula (\ref{thm3b}).

By using the relations $M(B+D)\leq M(B)+M(D)$, $m(B+D)\geq m(B)+m(D)$,
$m(B+D)\leq m(B)+M(D)$ for any $m\times m$ matrices $B$ and $D$, we have
$$\ba{rcl}
m(A)&=&
m(\sum_{i} \bar{B}_i\otimes \bar{C}_i + I_m\otimes \bar{C} +
\bar{B}\otimes I_n + S(A) I_m\otimes I_n)\\[3mm]
&\leq&m(I_m\otimes \bar{C} + \bar{B}\otimes I_n
+ S(A) I_m\otimes I_n)+M(\sum_{i} \bar{B}_i\otimes \bar{C}_i)\\[3mm]
&=&m(B)+m(C)+S(A)+M(\sum_{i} \bar{B}_i\otimes \bar{C}_i)\\[3mm]
&\leq& S(A) + \sum_{i} M(\bar{B}_i)M(\bar{C}_i).
\ea
$$
Hence formula (\ref{thm3c}) follows. \hfill $\square$

Generally $q_{B,C}$ with respect to our decomposition
$A=\sum_{i=1}^r B_i\otimes C_i$ does not equal to the separability indicator
$S(A)$. Suppose we have another decomposition
$A=\sum_{i=1}^{r^\prime} B_i^\prime\otimes C_i^\prime$.
As $B_i$ (and $C_i$) are defined in terms of the singular value decomposition
eigenvectors, they are linear independent. We can choose linear functionals
$\varphi_i$ such that $\varphi_i(B_j)=\delta_{ij}$. Applying $\varphi_i\otimes 1$ to both
sides of $\sum_{i=1}^r B_i\otimes C_i=\sum_{i=1}^{r^\prime} B_i^\prime\otimes C_i^\prime$
we get $C_i=\sum_{i=1}^{r^\prime}\varphi_i(B_j^\prime)C_j^\prime$, i.e.,
$C_i\in\, <C_1^\prime,...,C_{r^\prime}^\prime>$. Similarly we have
$B_i\in\, <B_1^\prime,...,B_{r^\prime}^\prime>$. Therefore any other Hermitian decomposition
$A=\sum_{i=1}^{r^\prime} B_i^\prime\otimes C_i^\prime$ can be obtained from our decomposition
$A=\sum_{i=1}^{r} B_i\otimes C_i$ in terms of the following transformations
$$
B_j^\prime=\sum_{i=1}^r E_{ij} B_i,~~~~
C_j^\prime=\sum_{i=1}^r F_{ij} C_i,
$$
as long as the real matrices $E=(E_{ij})$ and $F=(F_{ij})$ satisfy the relation $EF^t=I_r$.

The inequalities (\ref{thm3a})-(\ref{thm3c}) can be served as separability criterion themselves.
For instance, if the minimum eigenvalue of $A$ is zero, then $A$ is entangled if the right
hand side of (\ref{thm3b}) is great than zero.

\section{Conclusion and remarks}

We have developed a method of Hermitian tensor product
approximation of general complex matrices. From which an explicit
construction of density matrices on $H_1\otimes H_2$
in terms of the sum of tensor products of Hermitian matrices
on $H_1$ and $H_2$ is presented. From this construction we
have shown that a state is separable if and only if the separability
indicator is positive. In principle one can always get a
Hermitian tensor product decomposition of a density matrix
by using a basic set of Hermitian matrices. Our approach gives
a decomposition with minimum terms (the number of the terms depends on the
rank of $\hat{A}_{22}$), similar to the Schmidt decomposition
for bipartite pure states. In example (\ref{rhobf}) we see that
the $8\times 8$ density matrix $\rho_b$ has only $4$ terms in the tensor
product decomposition.

In \cite{vidaltarrach} an entanglement measure called robustness is introduced.
For a mixed state $\rho$ and a separable state $\rho_s$, the robustness of $\rho$
relative to $\rho_s$, $R(\rho||\rho_s)$, is the minimal $s \geq 0$ for which the
density matrix $(\rho + s\rho_s)/(1+s)$ is separable, i.e. the minimal
amount of mixing with locally prepared states which washes out all entanglement.
In particular, the random robustness of $\rho$ is the one when $\rho_s$ is taken to
be the (separable) identity matrix. In this case $\rho$ has the form
$\rho=(1+t)\rho_s^+ - t\,I_m\otimes I_n/mn)$, where $\rho_s^+$ is
separable. $\rho$ is separable if and if the minimum of $t$ is zero.
Therefore the separability indicator appeared from our matrix decompositions
is basically the minus of the random robustness, up to a normalization.
Another interesting separable approximations of density matrices was presented
in \cite{karnas}.  This method, so called best separable approximations, was based
on subtracting projections on product vectors from a
given density matrix in such a way that the remainder
remained positively defined. In stead expressing a density matrix
as the sum of a separable part and the identity part, this approximation
gives rise to a sum of a separable part and
an entangled part from which no more projections on product vectors
can be subtracted.

The results can be generalized to multipartite states.
Let's consider a general $l$-partite mixed state
$\rho_{1,2,...,l}$ on space $H_1\otimes H_2\otimes...\otimes H_l$.
We can first consider $\rho_{1,2,...,l}$ as a bipartite
state on space $H_1$ and $H_2\otimes...\otimes H_l$. By using Theorem 2 we
can find the tensor decomposition
$\rho_{1,2,...,l}=\sum_{i=1}^{r_1} B_i^1\otimes B_i^{23...l}$, where
$B_i^1$ and $B_i^{23...l}$ are Hermitian matrices on
$H_1$ and $H_2\otimes...\otimes H_l$ respectively.
The matrices $B_i^{23...l}$ can be again decomposed as
$B_i^{23...l}=\sum_{j=1}^{r_2} B_{ij}^2\otimes B_{ij}^{3...l}$, $\forall\, i$,
with $B_{ij}^2$ and $B_{ij}^{3...l}$ being Hermitian matrices
on $H_2$ and $H_3\otimes...\otimes H_l$ respectively. In doing so
at last we have the Hermitian tensor product decomposition of the form,
$\rho_{1,2,...,l}=\sum_{i=1}^{r} B_i^1\otimes B_i^{2}\otimes...\otimes B_i^{l}$,
where $B_i^{k}$ are Hermitian matrices on $H_k$.
New decompositions can be obtained,
$\rho_{1,2,...,l}=\sum_{i=1}^{r^\prime} {B^\prime}_i^1\otimes {B^\prime}_i^{2}
\otimes...\otimes {B^\prime}_i^{l}$, where
${B^\prime}_j^k=\sum_{i=1}^r E_{ij}^k {B}_i^k$, $k=1,...,l$, $E^k=(E_{ij}^k)$ are the real
matrices satisfying $\sum_{j=1}^{r^\prime}E_{i_1j}^1 E_{i_2j}^2...E_{i_lj}^l=
\delta_{i_1i_2}\delta_{i_2i_3}...\delta_{i_{l-1}i_l}$.

For any given decompositions, in terms of the protocol (\ref{protocal}), one has
$\rho_{1,2,...,l}=\sum_{i=1}^{r^\prime} {B^\prime}_i^1\otimes {B^\prime}_i^{2}
\otimes...\otimes {B^\prime}_i^{l}+ q\, Id_1\otimes Id_2\otimes ...\otimes Id_l$,
where $Id_i$ is the identity matrix on $H_i$, (${B^\prime}_i^1$, ${B^\prime}_i^{2}$,
$...$, ${B^\prime}_i^{l}$) are
Hermitian matrices such that $m({B^\prime}_i^{k})=0$, or part of them (but
not all) are identity matrices.
The separability indicator $S(\rho_{1,2,...,l})$
is the maximal value of the parameter $q$ for all possible positive Hermitian
tensor product decompositions.
If the parameter $S(\rho_{1,2,...,l})\geq 0$, the state $\rho_{1,2,...,l}$ is separable,
otherwise it is entangled.

\vspace{0.5truecm}
\noindent {\bf Acknowledgments}\, S.M. Fei
gratefully acknowledges the warm hospitality of Dept. Math., NCSU and
the support provided by American Mathematical Society. Jing greatly acknowledges
the support from the NSA and Alexander von Humboldt foundation.


\begin{thebibliography}{99}
\bibitem{pre98} J. Preskill, The Theory of Quantum Information and Quantum
Computation, California Inst. of Tech., 2000,
http://www.theory.caltech.edu/people/preskill/ph229/.

\bibitem{nielsen} M.A. Nielsen and I.L. Chuang, Quantum Computation and
Quantum Information, Cambridge University Press, Cambridge, 2000.

\bibitem{zeilinger} D. Bouwmeester, A. Ekert and A. Zeilinger (Eds.), The
Physics of Quantum Information: Quantum Cryptography, Quantum Teleportation
and Quantum Computation, Springer, New York, 2000.

\bibitem{lbck00} M. Lewenstein, D. Bruss, J.I. Cirac, B. Kraus, M. Kus, J.
Samsonowicz, A. Sanpera and R. Tarrach, J. Mod. Opt. \textbf{47}, 2841 (2000).

\bibitem{terhal01} B.M. Terhal, Theor. Comput. Sci. \textbf{287}, 313 (2002).

\bibitem{3hreview} M. Horodecki, P. Horodecki and R. Horodecki, Springer
Tracts in Mod. Phy. \textbf{173}, 151 (2001).

\bibitem{Bell64} J.S. Bell, Physics (N.Y.) \textbf{1}, 195 (1964).

\bibitem{peres} A. Peres, Phys. Rev. Lett. \textbf{77}, 1413 (1996).

\bibitem{3hPLA223} M. Horodecki, P. Horodecki and R. Horodecki, Phys. Lett.
A \textbf{223}, 1 (1996).

\bibitem{2hPRA99} M. Horodecki and P. Horodecki, Phys. Rev. A \textbf{59}, 4206 (1999).

\bibitem{cag99} N.J. Cerf, C. Adami and R.M. Gingrich, Phys. Rev. A
\textbf{60}, 898 (1999).

\bibitem{nielson01} M.A. Nielsen and J. Kempe, Phys. Rev. Lett. \textbf{86}, 5184 (2001).

\bibitem{ter00} B. Terhal, Phys. Lett. A \textbf{271}, 319 (2000).

\bibitem{lkch00} M. Lewenstein, B. Kraus, J.I. Cirac and P. Horodecki, Phys.
Rev. A \textbf{62}, 052310 (2000).

\bibitem{ru02} O. Rudolph, Physical Review A \textbf{67}, 032312 (2003).

\bibitem{ChenQIC03} K. Chen and L.A. Wu, Quant. Inf. Comput. \textbf{3}, 193 (2003).

\bibitem{chenPLA02} K. Chen and L.A. Wu, Phys. Lett. A \textbf{306}, 14 (2002).

\bibitem{chenkai} S. Albeverio, K. Chen and S.M. Fei, Phys. Rev. A \textbf{68}, 062313 (2003).

\bibitem{hlvc00} P. Horodecki, M. Lewenstein, G. Vidal and I. Cirac, Phys.
Rev. A \textbf{62}, 032310 (2000).

\bibitem{afg01} S. Albeverio, S.M. Fei and D. Goswami, Phys. Lett. A \textbf{286},
 91 (2001).

\bibitem{feipla02} S.M. Fei, X.H. Gao, X.H. Wang, Z.X. Wang and K. Wu, Phys.
Lett. A \textbf{300}, 555 (2002).

\bibitem{loan} G.H. Golub, C.F. von Loan, {\it Matrix Computation},
The Johns Hopkins University Press, 1989.

\bibitem{loan1} C.F. Van Loan and N.P. Pitsianis, in: Linear Algebra for
Large Scale and Real Time Applications, M.S. Moonen and G.H. Golub (Eds.),
Kluwer Publications, 1993, pp. 293--314.

\bibitem{kropro} N.P. Pitsianis, Ph.D. thesis, {\it The Kronecker
Product in Approximation and Fast Transform Generation}, Cornell
University, New York (1997).

\bibitem{werner} R.F. Werner, Phys. Rev. A \textbf{40}, 4277 (1989).

\bibitem{horodecki97} P. Horodecki, Phys. Lett. A \textbf{232}, 333
(1997).

\bibitem{vidaltarrach} G. Vidal and R. Tarrach,
Phys.Rev. A \textbf{59}, 141(1999).

\bibitem{karnas} S. Karnas and M. Lewenstein,
J. Phys. A \textbf{34}, 6919(2001).

\end{thebibliography}
\end{document}